\documentclass[12pt]{article}

\usepackage{amsmath,amssymb,amsfonts,amsbsy}
\usepackage{cite}
\usepackage{graphicx}
\usepackage{wrapfig}
\usepackage{epsfig}

\usepackage{bbm} 
\usepackage{bm} 
\usepackage{color}                                                       %
\usepackage{dsfont} 
\usepackage{latexsym} 
\usepackage{lscape} 
\usepackage{mathrsfs} 
\usepackage{morefloats} 
\usepackage{floatflt} 
\usepackage{slashed} 
\usepackage{psfrag}

\DeclareFontFamily{OT1}{mygreek}{}%
\DeclareFontShape{OT1}{mygreek}{m}{n}{<->omsegr}{}%
\DeclareFontShape{OT1}{mygreek}{b}{n}{<->omsegrb}{}%
\DeclareFontShape{OT1}{mygreek}{m}{it}{<->omsegri}{}%
\DeclareFontShape{OT1}{mygreek}{bx}{n}{<->sub * mygreek/b/n}{}%
\DeclareFontShape{OT1}{mygreek}{m}{sl}{<->sub * mygreek/m/it}{}%
\DeclareSymbolFont{Greekrm}{OT1}{mygreek}{m}{n}
\DeclareSymbolFont{Greekbf}{OT1}{mygreek}{b}{n}
\DeclareSymbolFont{Greekit}{OT1}{mygreek}{m}{it}
\DeclareMathSymbol{\omegab}{\mathalpha}{Greekbf}{119}

\begin{document}
\addcontentsline{toc}{subsection}{{ON THE DISPUTED $\pi_1(1600)$ RESONANCE AND OBSERVATION OF A NEW ISO-VECTOR RESONANCE}\\
{\it F. Nerling}}
\graphicspath{{exper/Frank Nerling/}}

\setcounter{section}{0}
\setcounter{subsection}{0}
\setcounter{equation}{0}
\setcounter{figure}{0}
\setcounter{footnote}{0}
\setcounter{table}{0}

\begin{center}
\textbf{ON THE DISPUTED $\pi_1(1600)$ RESONANCE AND OBSERVATION OF A NEW ISO-VECTOR RESONANCE}

\vspace{5mm}

F.~Nerling$^{\,1\,\dag}$ \\on behalf of the COMPASS collaboration

\vspace{5mm}

\begin{small}
  (1) \emph{Institut f\"ur Kernphysik, Universit\"at Mainz, Becherweg 45, 55099 Mainz, Germany} \\
  $\dag$ \emph{E-mail: nerling@cern.ch}
\end{small}
\end{center}

\vspace{0.0mm} 

\begin{abstract}
The COMPASS experiment at CERN delivers new results on the search for exotic mesons. 
A spin-exotic resonance, the $\pi_1(1600)$, was reported by several experiments in the past. 
Those observations are, however, still to date highly disputed in the community.
Especially the $\rho\pi$ decay channel allows for simultaneous observation of well established 
and less known resonances in different decay modes. 
The results from amplitude analysis of diffractively produced $(3\pi)^{-}$ final states 
show consistently a spin-exotic signal, that appears in agreement with
previous observations of the $\pi_1(1600)$. The high-statistics 2008 data sample allows and demands 
for an extended amplitude analysis method that further disentangles resonant and non-resonant particle production. 
The present status of analysis of COMPASS data and the observation of a new iso-vector meson 
$a_1(1420)$ is discussed.
\end{abstract}

\paragraph{Introduction}~\\
Exotic mesons have been reported by different experiments and in different decay channels. 
Quantum Chromodynamics (QCD) allows for and predicts exotic mesons like glue-balls, hybrids or 
tetraquarks according to several models. The experimental observation of so-called spin-exotic 
mesons, like the $\pi_1(1600)$ having exotic $J^{PC}$ quantum numbers not accessible 
within the naive Constituent Quark Model, would be a fundamental confirmation of QCD, for a 
recent overview see e.g.~\cite{MeyerHaarlem2010}.
Especially the resonant nature of signals observed in the exotic $J^{PC}=1^{-+}$ partial-wave 
of the $\rho\pi$ decay channel, accessible via 3$\pi$ final states, as reported by the E852/BNL and 
the VES experiments~\cite{Adams:1998,Khokhlov:2000} in $\pi^{-}\pi^{+}\pi^{-}$ final states are 
questioned. The conclusions were withdrawn in later publications~\cite{Amelin:2005} and re-analyses 
of the $(3\pi)^{-}$ system in two decay modes (charged: $\pi^{-}\pi^{+}\pi^{-}$ and neutral: $\pi^{-}\pi^{0}\pi^{0}$) led to opposite 
conclusions~\cite{Dzierba:2006}. One may get a hint at this controversy looking at~\cite{PDG}. 

The data taken with the COMPASS experiment at the CERN SPS provide excellent opportunity for the 
search for exotic resonances. In the 2004 pilot run data ($\pi^{-}$ beam, Pb target), a significant 
$J^{PC}$ spin-exotic signal at $1660$$\pm$$10^{+0}_{-64}$\,MeV/c$^2$ is observed that is consistent with 
the disputed $\pi_1(1600)$; it shows a clean phase motion against well-known resonances~\cite{Alekseev:2009a}. 
The high statistics of the 2008 proton target data allows the search for exotic states in different decay 
modes in the same experiment, cf.~\cite{nerling:2009}. Employing the same PWA analysis method as 
in~\cite{Alekseev:2009a}, the results obtained for the $J^{PC}M^\epsilon$=$(1^{-+})1^{+}\rho^{-}\pi^{0}$ 
and $(1^{-+})1^{+}\rho^0\pi^{-}$ intensity and relative phase
are similar to the previous observations~\cite{nerling:2012b,haas:2011}. 
Apart of the established resonances $a_1(1260)$, $a_2(1320)$, $\pi_2(1670)$, also $\pi(1800)$ and $a_4(2040)$, 
an exotic signal in the $1^{-+}$ wave at around 1.6\,GeV/$c^2$ is observed, that shows a clean phase motion with 
respect to well-known resonances. 
These results are consistently obtained for both $\rho\pi$ decay modes, neutral and charged.
%
\begin{figure}[tp!]
  \begin{minipage}[h]{.32\textwidth}
    \begin{center}
         \vspace{-0.5cm}
\resizebox{1.0\columnwidth}{!}{%
  \includegraphics[clip,trim= 10 5 30 25, width=1.1\linewidth, angle=0]{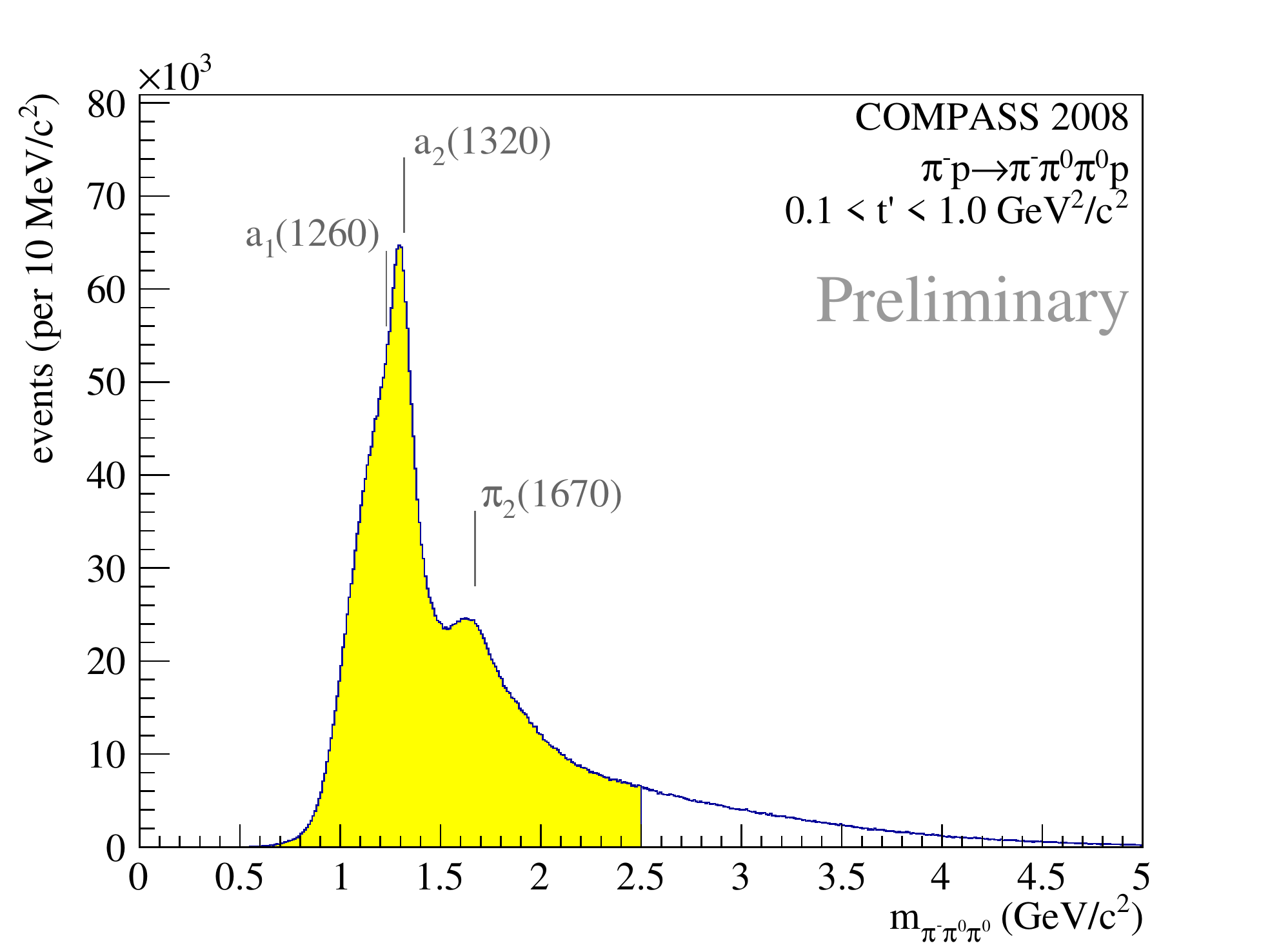} }
    \end{center}
  \end{minipage}
  \hfill
  \begin{minipage}[h]{.32\textwidth}
    \begin{center}
      \vspace{-0.5cm}
\resizebox{1.0\columnwidth}{!}{%
     \includegraphics[clip,trim= 10 5 30 25, width=1.1\linewidth, angle=0]{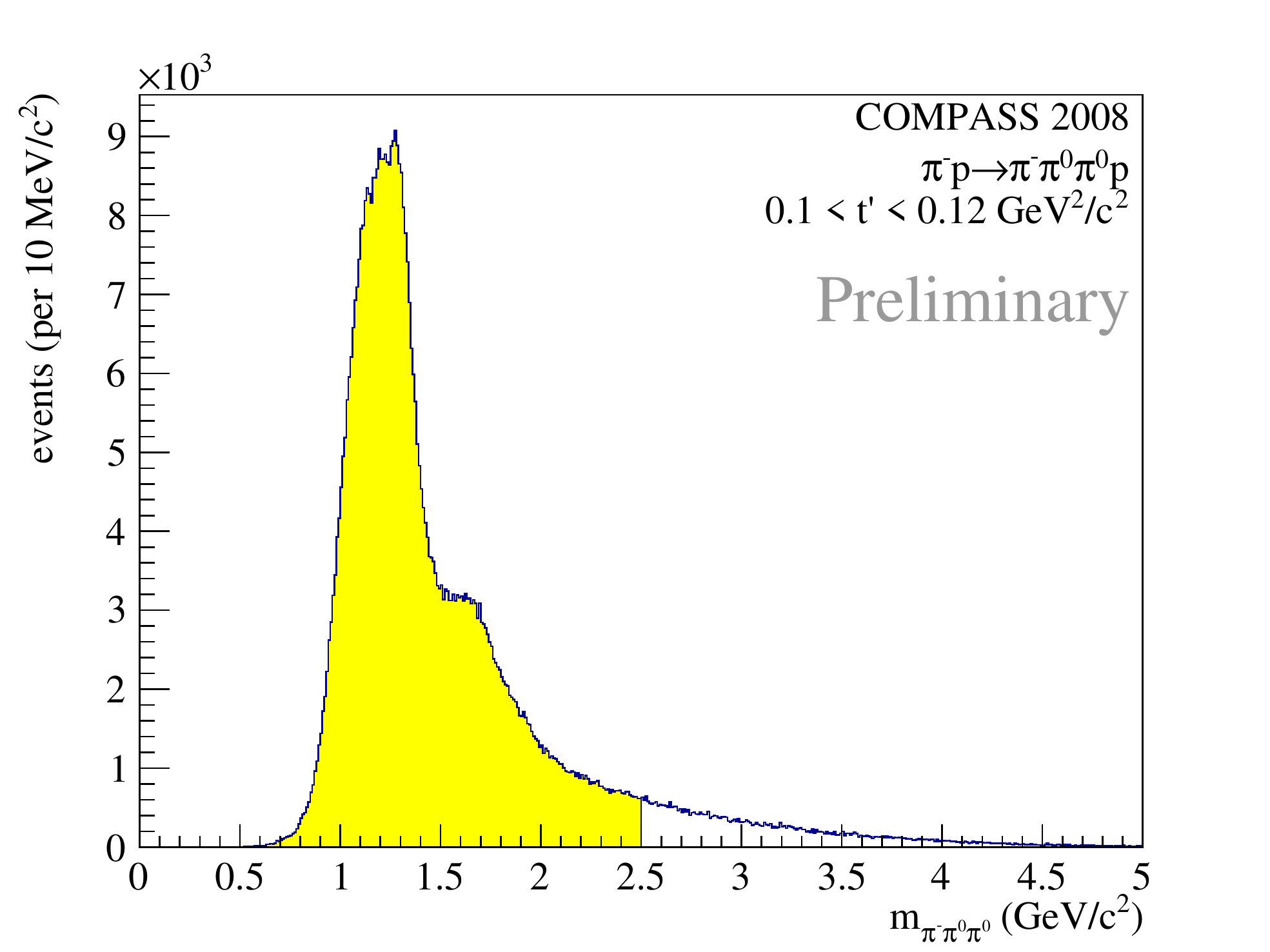} }
    \end{center}
  \end{minipage}
  \begin{minipage}[h]{.32\textwidth}
    \begin{center}
      \vspace{-0.5cm}
\resizebox{1.0\columnwidth}{!}{%
     \includegraphics[clip,trim= 10 5 30 25, width=1.1\linewidth, angle=0]{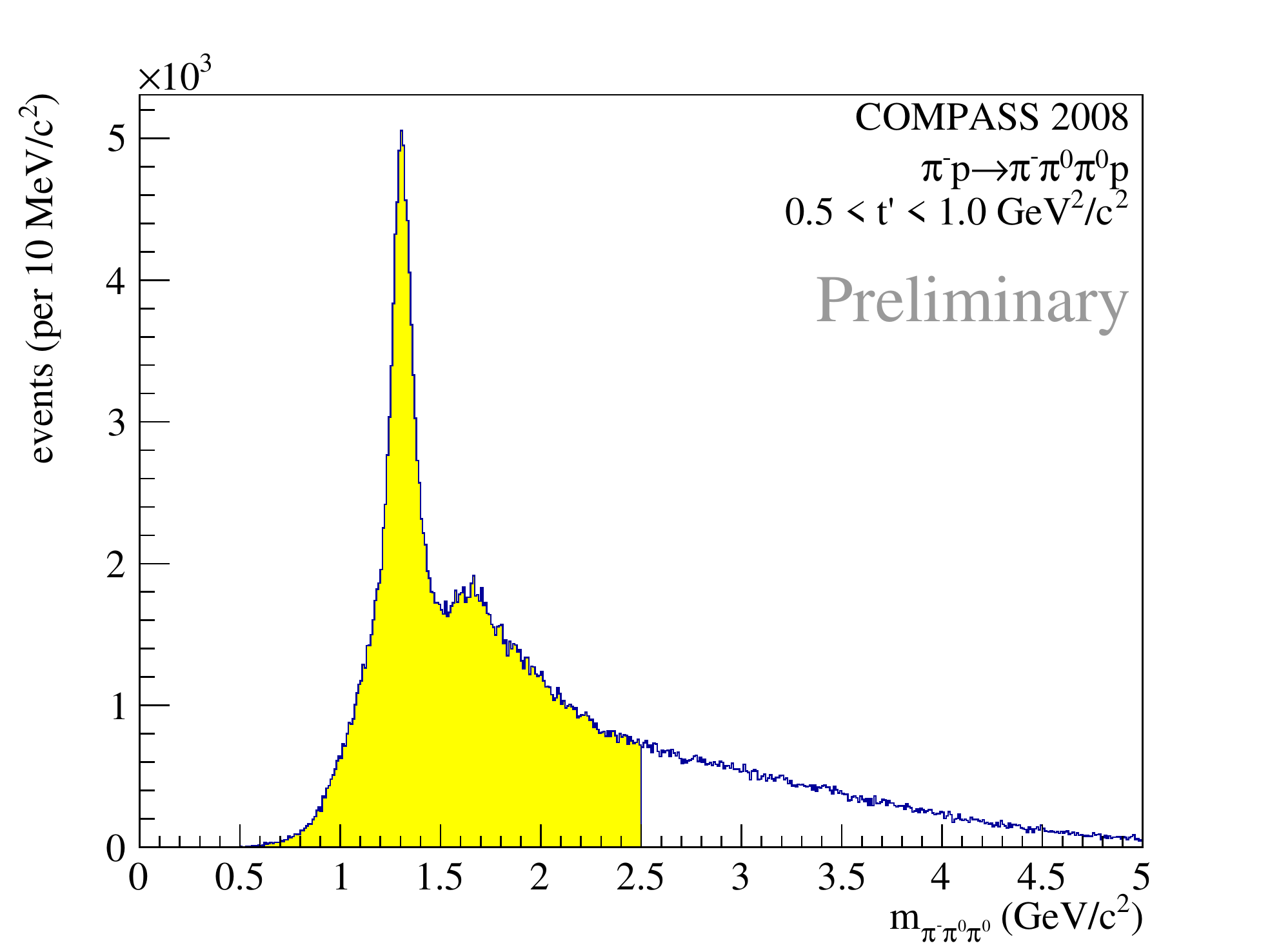}}
    \end{center}
  \end{minipage}
   \begin{minipage}[h]{1.0\textwidth}
     \begin{center}
      \vspace{+0.2cm}
 \resizebox{1.0\columnwidth}{!}{%
   \includegraphics[clip,trim= 0 0 0 0, width=0.4\linewidth, angle=0]{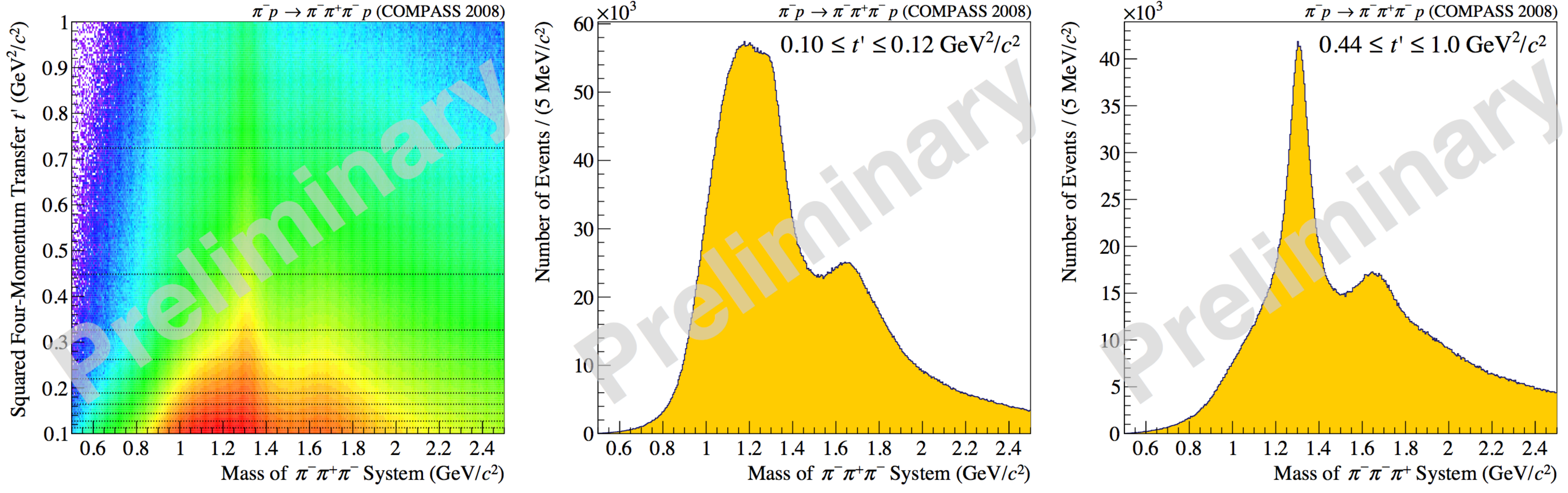} }
     \end{center}
   \end{minipage}
    \begin{center}
      \vspace{-0.5cm}
     \caption{Mass spectra of the $(3\pi)^{-}$ systems. \textit{Top: Neutral mode ---} Different regions of momentum transfer $t'$: 
       ``whole $t'$'' \textit{(left)}, ``low $t'$'' \textit{(centre)} and ``high $t'$''\textit{(right)}. 
       \textit{Bottom: Charged mode ---} Correlation of $m_{\rm 3\pi}$ and $t'$ \textit{(left)}, ``low $t'$'' \textit{(centre)} and 
       ``high $t'$'' \textit{(right)}}
       \label{fig:3piMassTot_neutral_charged}
     \end{center}
     \vspace{-0.7cm}
\end{figure}
\paragraph{New Partial-Wave Analysis Results}~\\
The data analysed for both $(3\pi)^-$ decay modes has been extended to the full 2008 proton target data.
The total outgoing $3\pi$-mass spectrum is shown for the neutral mode data in Fig.\,\ref{fig:3piMassTot_neutral_charged} 
(top) for the whole range of momentum transfer $t'$ (Fig.\,\ref{fig:3piMassTot_neutral_charged}, top, left), and for ranges 
of ``low'' and ``high'' (Fig.\,\ref{fig:3piMassTot_neutral_charged}, top, centre and right) $t'$ values, indicating a dependence 
of $m_{\rm 3\pi}$ on $t'$.
This dependence is shown for the charged data (Fig.\,\ref{fig:3piMassTot_neutral_charged}, bottom) at its whole glance in terms 
of a 2D plot of $t'$ vs. $m_{\rm 3\pi}$ (Fig.\,\ref{fig:3piMassTot_neutral_charged}, bottom, left). The two exemplary $m_{\rm 3\pi}$ 
spectra for the ``low'' and ``high'' $t'$ values similar to the ones shown for the neutral mode 
(Fig.\,\ref{fig:3piMassTot_neutral_charged}, top, centre and right) are given 
for comparison for the charged mode (Fig.\,\ref{fig:3piMassTot_neutral_charged}, bottom, centre and right) 
as well, illustrating the similarity of the $t'$ dependence observed for both, neutral and charged mode data. 

Given the observed dependence on $t'$ and the large statistics ($\sim$50\,M $\pi^-\pi^+\pi^-$, $\sim$3.5\,M 
$\pi^-\pi^0\pi^0$ events), the partial-wave analysis (PWA) method has been extended w.r.t. to the previous scheme 
of a two-step PWA as applied previously~\cite{Alekseev:2009a,nerling:2012b}. The first step analysis, the mass-independent 
PWA, has now been performed in different ranges of $t'$ (a scheme already addressed in~\cite{Accmor:1980,Dzierba:2006}) with 
equal statistics contained in each bin, which is then completed by the second step mass-dependent Breit-Wigner $\chi^2$ fit, 
that takes into account the observed $t'$ dependencies by performing a simultaneous optimisation of the resonant parameters 
in all $t'$ regions. 
This procedure disentangles resonant from 
non-resonant particle production (e.g. dynamically produced components caused by the Deck effect~\cite{Deck:1964}), the 
former should not depend on $t'$, whereas non-resonant backgrounds may well do so. 

For the results presented here, the data has been divided into 8 and 11 bins of $t'$ for the neutral and charged mode data, 
respectively, as illustrated for the charged data in Fig.\,\ref{fig:3piMassTot_neutral_charged} (bottom, left). 
The mass-independent PWA has been performed for each of the 8 and 11 ranges of $t'$, whereas 40\,MeV/$c^2$ (neutral mode) and 
20\,MeV/$c^2$ (charged mode) wide $m_{\rm 3\pi}$ bins have been chosen as previously~\cite{nerling:2012b,haas:2011}.
The wave-set has been extended from 53 to 88 partial-waves. The mass-dependent PWA including 6 out of the 
88 waves is released for public merely for the charged mode data, see also \cite{paul:2013,suhl:2013} for more details. 
%
%
\begin{figure}[tp!]
  \begin{minipage}[h]{.32\textwidth}
    \begin{center}
         \vspace{-0.5cm}
\resizebox{1.0\columnwidth}{!}{%
  \includegraphics[clip,trim= 5 0 10 15, width=1.0\linewidth, angle=0]{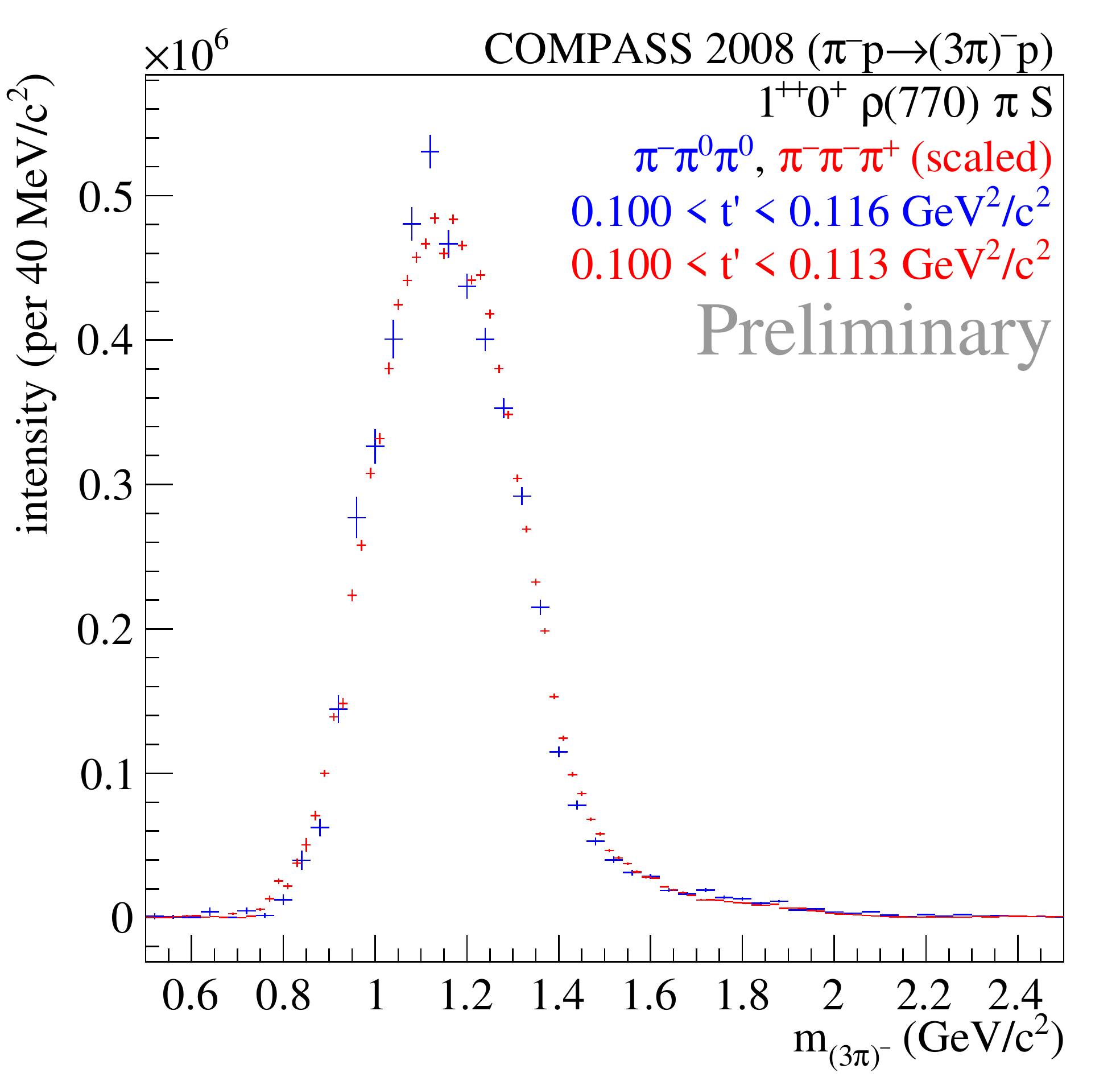} }
    \end{center}
  \end{minipage}
  \hfill
  \begin{minipage}[h]{.32\textwidth}
    \begin{center}
      \vspace{-0.5cm}
\resizebox{1.0\columnwidth}{!}{%
     \includegraphics[clip,trim= 5 0 10 15, width=1.0\linewidth, angle=0]{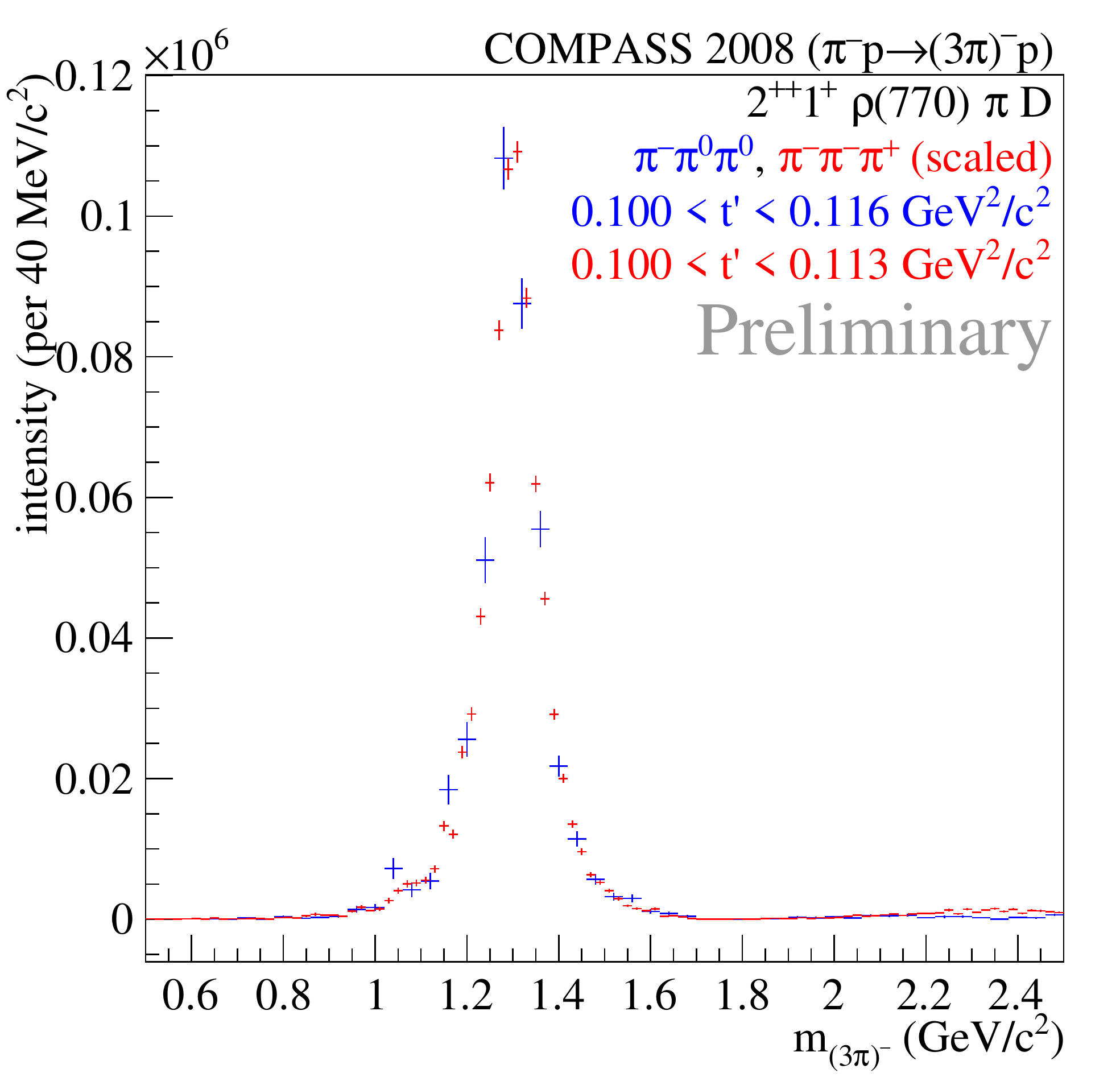} }
    \end{center}
  \end{minipage}
  \begin{minipage}[h]{.32\textwidth}
    \begin{center}
      \vspace{-0.5cm}
\resizebox{1.0\columnwidth}{!}{%
     \includegraphics[clip,trim= 5 -5 10 20, width=1.0\linewidth, angle=0]{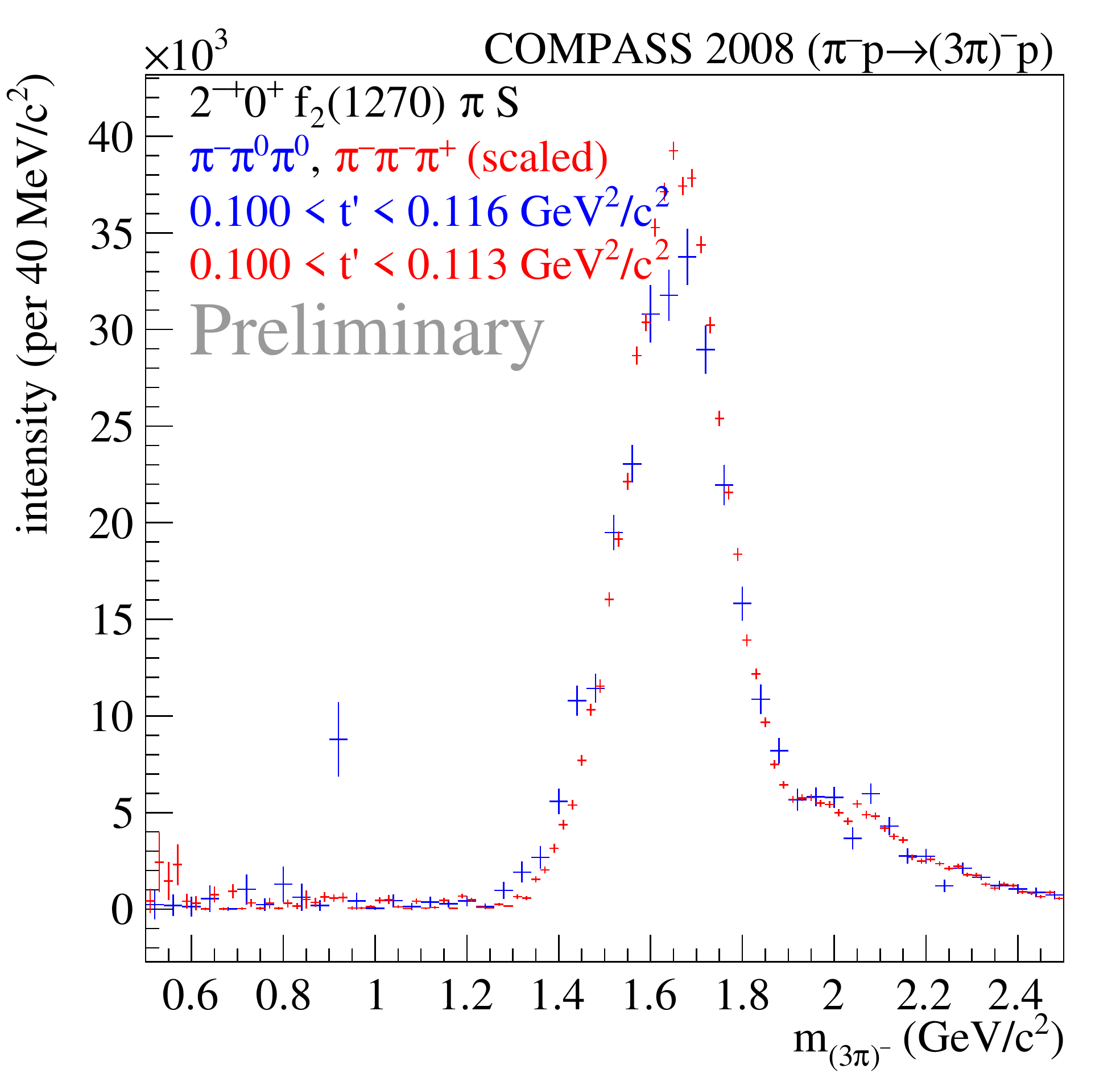}}
    \end{center}
  \end{minipage}
  \begin{minipage}[h]{.32\textwidth}
    \begin{center}
      \vspace{-0.2cm}
\resizebox{1.0\columnwidth}{!}{%
  \includegraphics[clip,trim= 5 0 10 15, width=1.0\linewidth, angle=0]{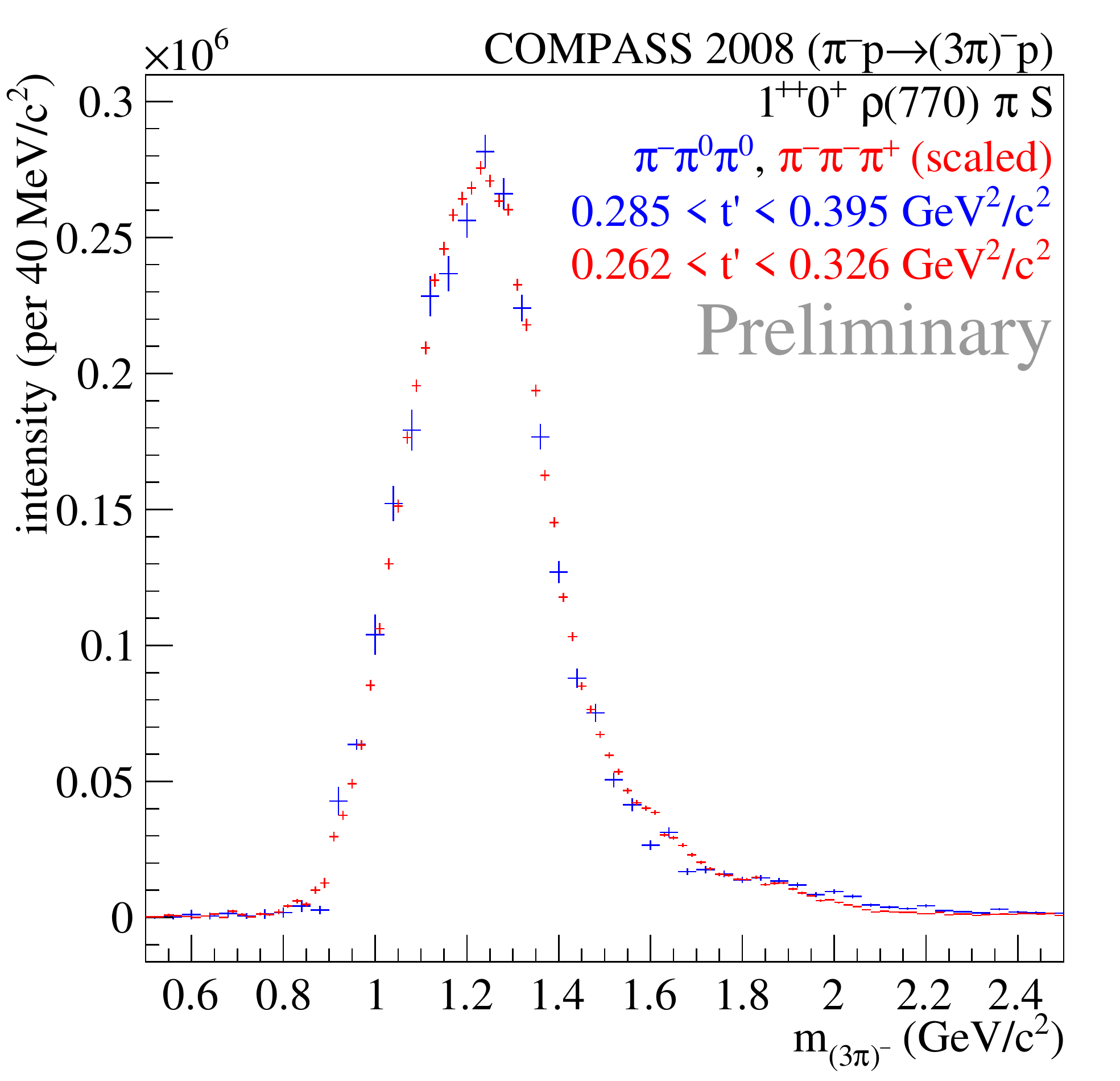} }
    \end{center}
  \end{minipage}
  \hfill
  \begin{minipage}[h]{.32\textwidth}
    \begin{center}
      \vspace{-0.2cm}
\resizebox{1.0\columnwidth}{!}{%
     \includegraphics[clip,trim= 5 0 10 15, width=1.0\linewidth, angle=0]{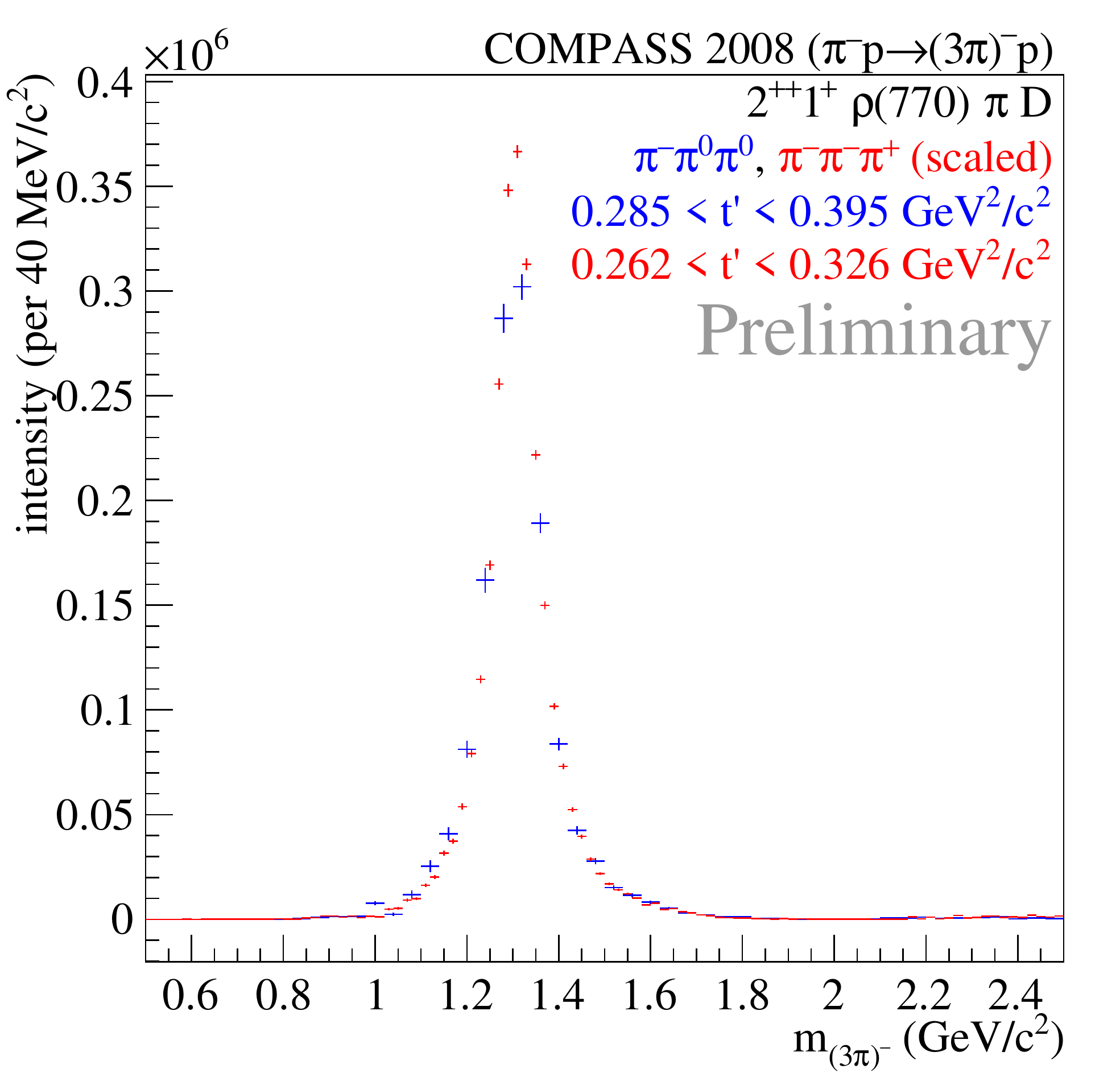} }
    \end{center}
  \end{minipage}
  \begin{minipage}[h]{.32\textwidth}
    \begin{center}
      \vspace{-0.1cm}
\resizebox{1.0\columnwidth}{!}{%
     \includegraphics[clip,trim= 5 0 10 15, width=1.0\linewidth, angle=0]{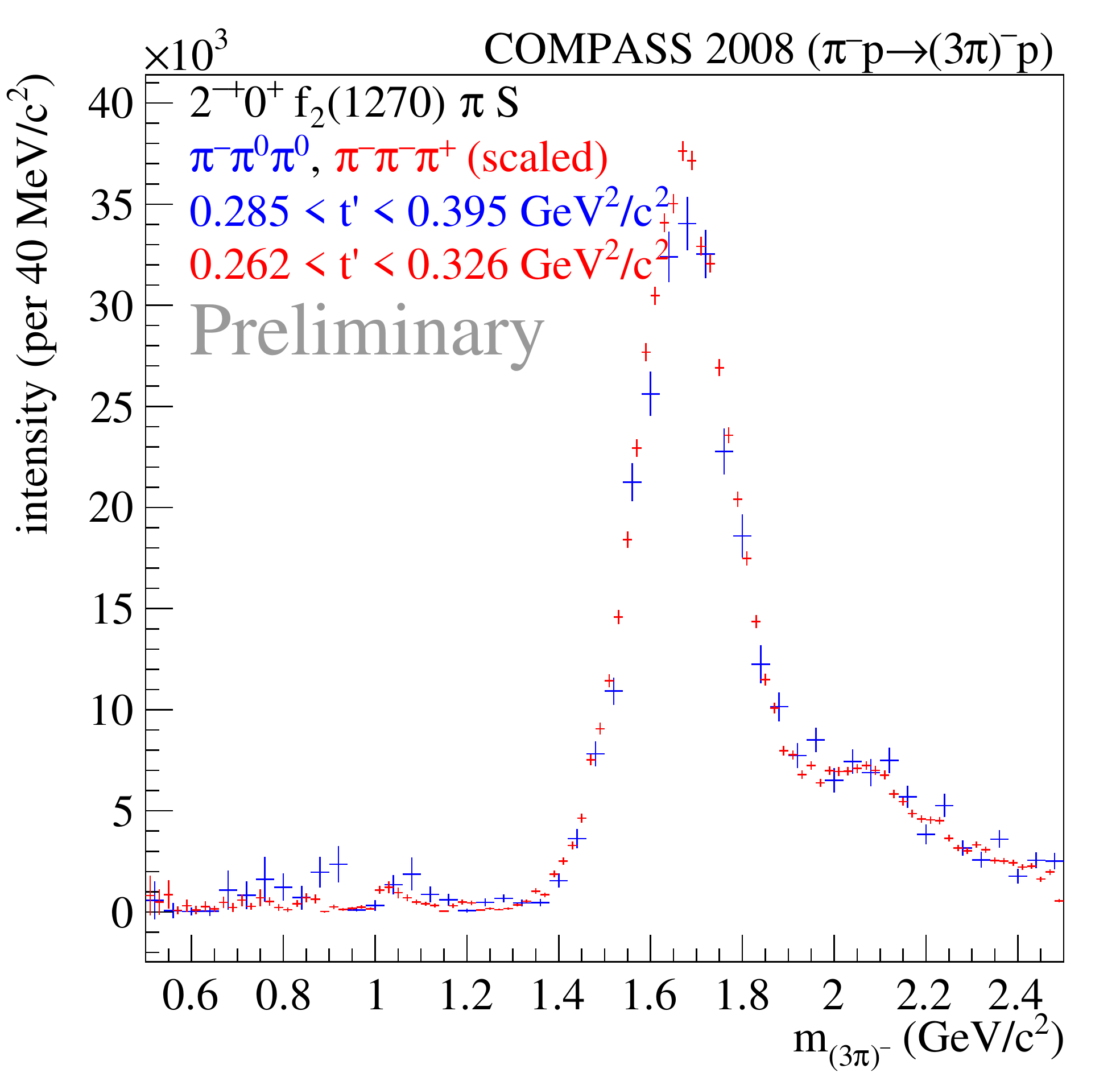}}
    \end{center}
  \end{minipage}
    \begin{center}
      \vspace{-0.7cm}
     \caption{Mass-independent PWA result for neutral (blue) vs. charged (red) mode shown for the major waves for 
       ``low'' {\it (top)} and ``high'' {\it (bottom)} values of $t'$.  
       The $a_1(1260)$ in the $(1^{++})0^+\rho(770)\pi~S$-wave {\it (left)}, the $a_2(1320)$ in the 
       $(2^{++})1^{+}\rho(770)\pi~D$-wave {\it (centre)} and the $\pi_2(1670)$ in the
       $(2^{-+})0^+f_2(1270)\pi~S$-wave {\it (right)} are shown.}
       \label{fig:MajorWavesTprimeDep}
     \end{center}
     \vspace{-0.7cm}
\end{figure}

\begin{figure}[tp!]
    \begin{center}
         \vspace{-0.5cm}
         \resizebox{1.0\columnwidth}{!}{%
  \includegraphics[clip,trim= 0 0 0 0, width=1.0\linewidth, angle=0]{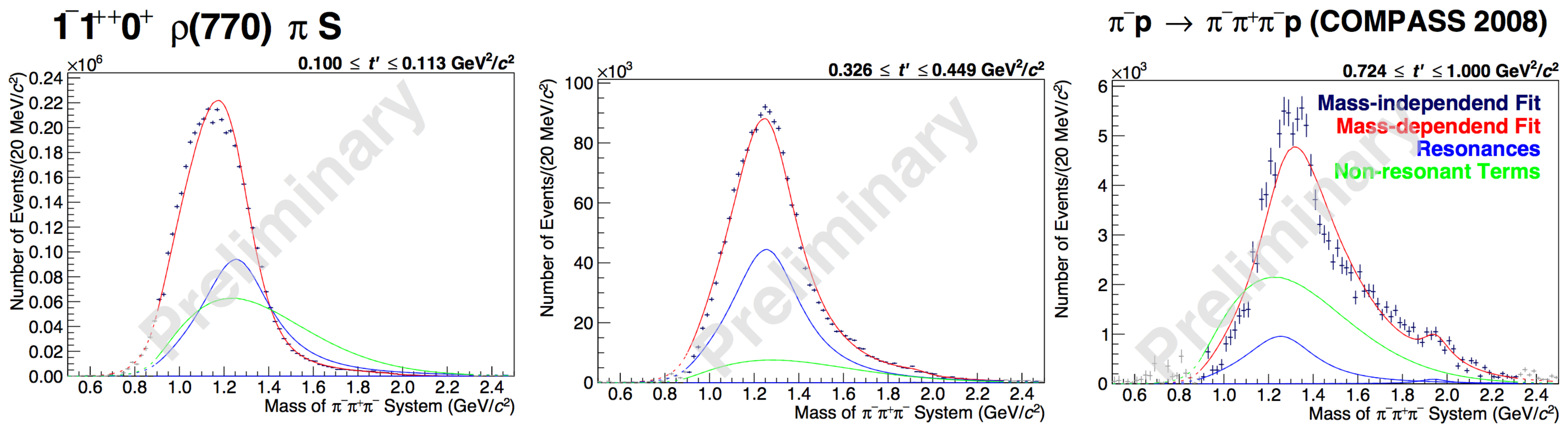} }
      \vspace{-0.7cm}
     \caption{Complete PWA result for the $(1^{++})0^+\rho(770)\pi~S$-wave, in which we observe the $a_1(1260)$, 
       shown for three exemplary bins in $t'$ (``low'', ``medium'', and ``high''). Shown are the mass-independent 
       fit result (data points) and overlaid the fitted Breit-Wigner (BW) description (red curve), which consists of the 
       BW describing the $a_1(1260)$ (blue curve) and the non-resonant background contribution (green curve).
     }
       \label{fig:MassDep3pic_a1_3bins}
     \end{center}
     \vspace{-0.7cm}
\end{figure}

In Fig.\,\ref{fig:MajorWavesTprimeDep}, the mass-independently fitted intensities for the major waves are shown for 
exemplary two very different ranges of $t'$  (``low'' and ``high'') for both, the neutral and the charged mode fit results
(normalised to the integral for each plot to compare the shapes). The main resonances $a_1(1260)$ (Fig.\,\ref{fig:MajorWavesTprimeDep}, 
left), $a_2(1320)$ (Fig.\,\ref{fig:MajorWavesTprimeDep}, centre), and $\pi_2(1670)$ (Fig.\,\ref{fig:MajorWavesTprimeDep}, right) are 
consistently observed for the neutral vs. charged mode data, the shapes are mostly coinciding for the different $t'$ ranges ``low'' and 
``high'' (Fig.\,\ref{fig:MajorWavesTprimeDep}, top and bottom). 
While the $a_2(1320)$ and $\pi_2(1670)$  (Fig.\,\ref{fig:MajorWavesTprimeDep}, centre and right) 
are observed rather robust in shape against $t'$, the $a_1(1260)$ (Fig.\,\ref{fig:MajorWavesTprimeDep}, left) shows a significant shift 
in mass for the different $t'$. In addition, structures around the $a_1(1260)$ and $\pi_2(1670)$ 
(Fig.\,\ref{fig:MajorWavesTprimeDep}, left and right) reveal underlying dynamics resulting in a $t'$ dependent shape, which cannot 
be solely attributed to a resonance. This will be resolved by the completing second step, the mass-dependent fit. Fitting the Breit-Wigner  
description of the resonances simultaneously to the different $t'$ ranges, while allowing for background contributions, indeed 
different relative contributions of non-resonant background are found, depending on $t'$ (Fig.\,\ref{fig:MassDep3pic_a1_3bins}). 
%
%

\begin{figure}[bp!]
  \begin{minipage}[h]{.32\textwidth}
    \begin{center}
         \vspace{-0.5cm}
\resizebox{1.0\columnwidth}{!}{%
  \includegraphics[clip,trim= 5 0 10 15, width=1.0\linewidth, angle=0]{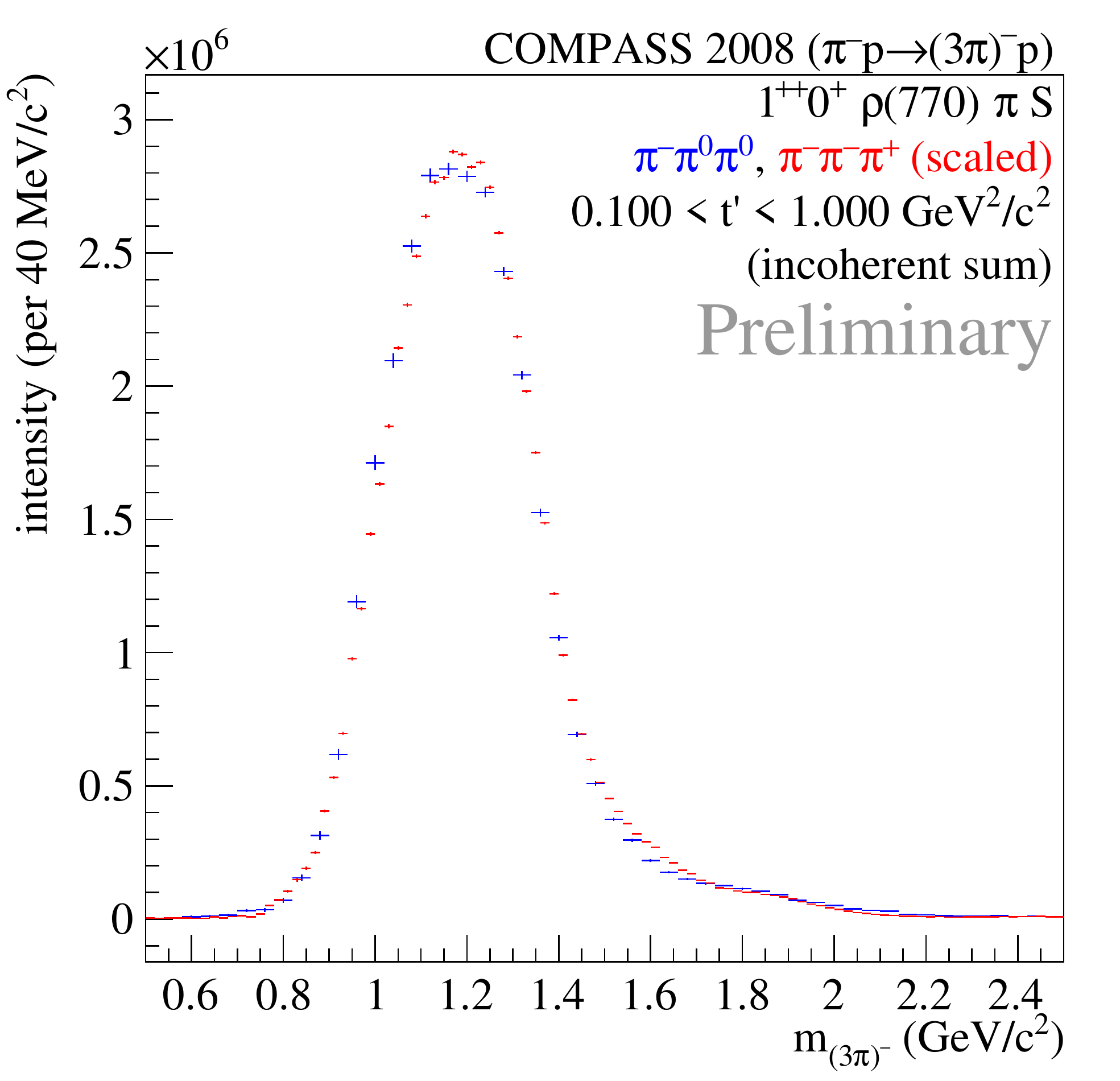} }
    \end{center}
  \end{minipage}
  \hfill
  \begin{minipage}[h]{.32\textwidth}
    \begin{center}
      \vspace{-0.5cm}
\resizebox{1.0\columnwidth}{!}{%
     \includegraphics[clip,trim= 5 0 10 15, width=1.0\linewidth, angle=0]{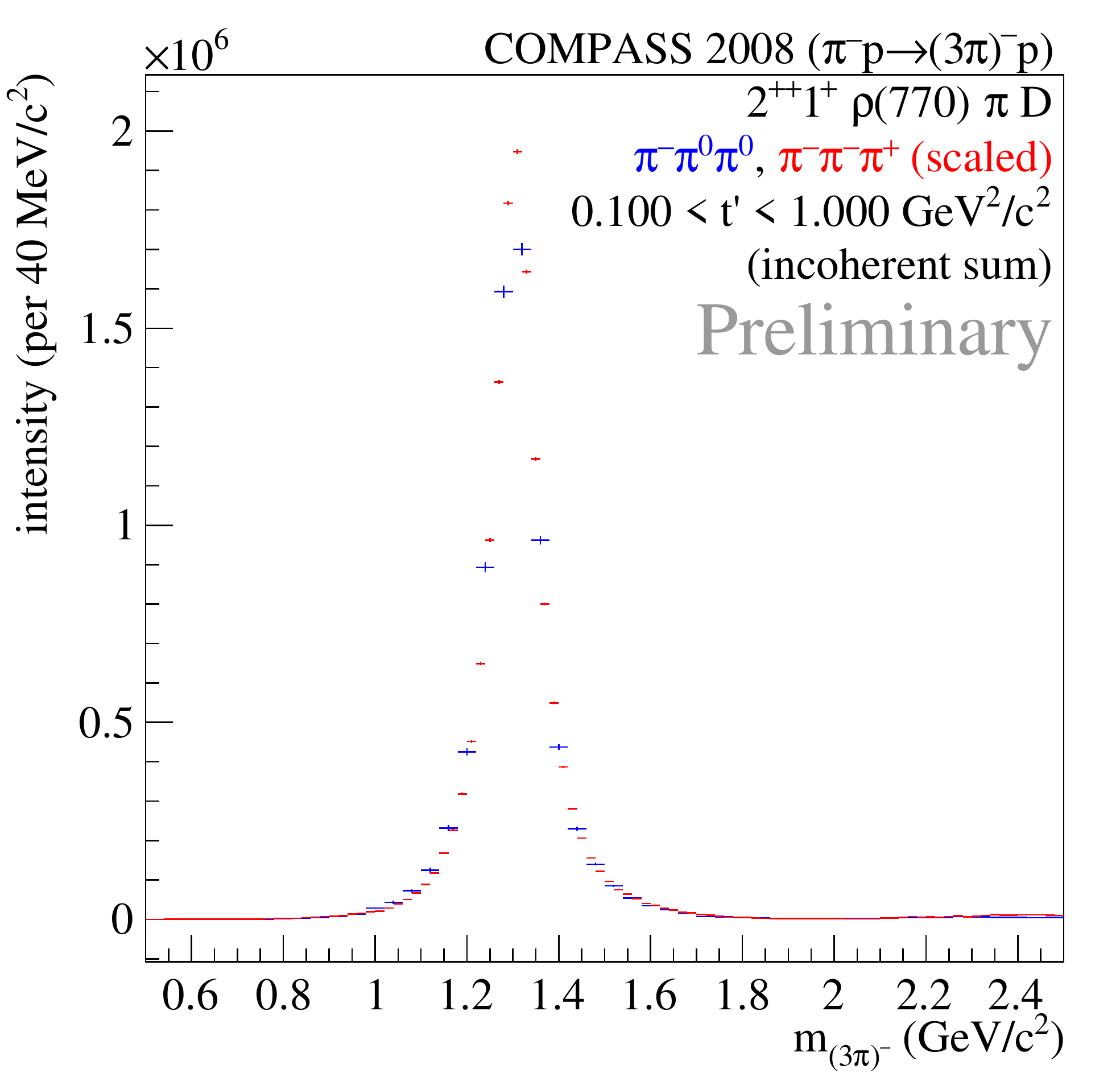} }
    \end{center}
  \end{minipage}
  \begin{minipage}[h]{.32\textwidth}
    \begin{center}
      \vspace{-0.5cm}
\resizebox{1.0\columnwidth}{!}{%
     \includegraphics[clip,trim= 5 -5 10 20, width=1.0\linewidth, angle=0]{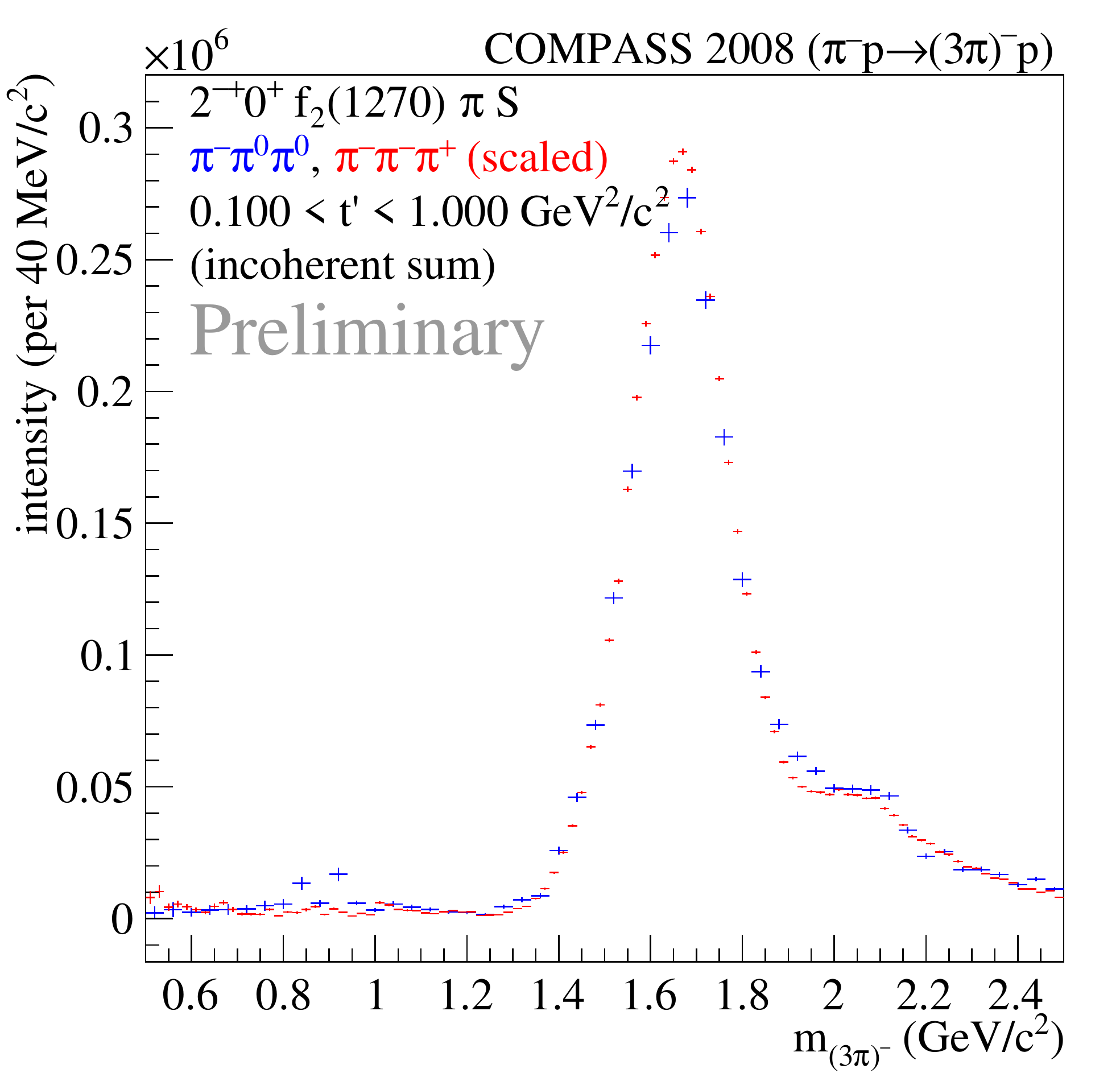}}
    \end{center}
  \end{minipage}
  \begin{minipage}[h]{.32\textwidth}
    \begin{center}
      \vspace{-0.2cm}
\resizebox{1.0\columnwidth}{!}{%
  \includegraphics[clip,trim= 5 0 10 15, width=1.0\linewidth, angle=0]{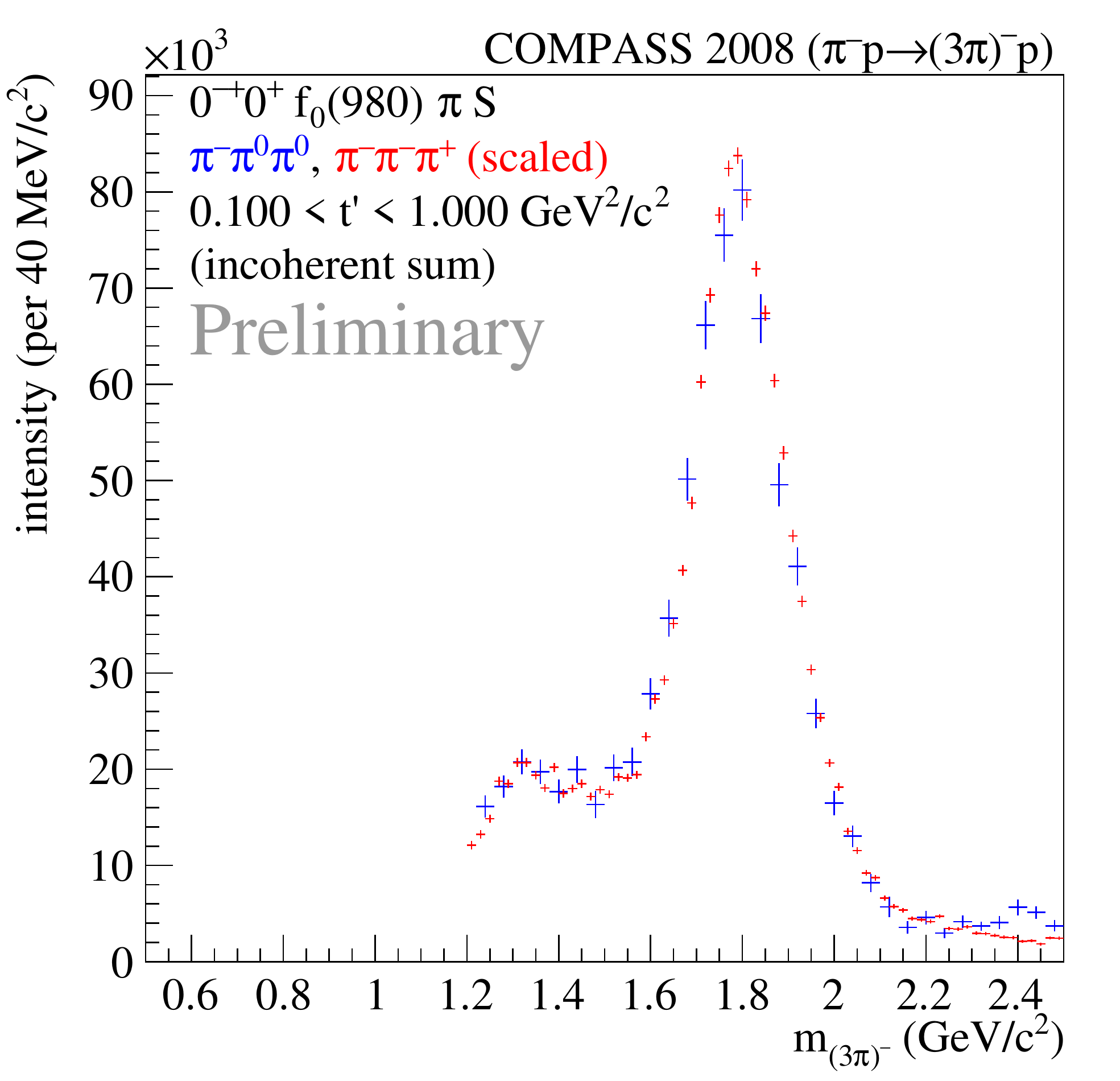} }
    \end{center}
  \end{minipage}
  \hfill
  \begin{minipage}[h]{.32\textwidth}
    \begin{center}
      \vspace{-0.2cm}
\resizebox{1.0\columnwidth}{!}{%
     \includegraphics[clip,trim= 5 0 10 15, width=1.0\linewidth, angle=0]{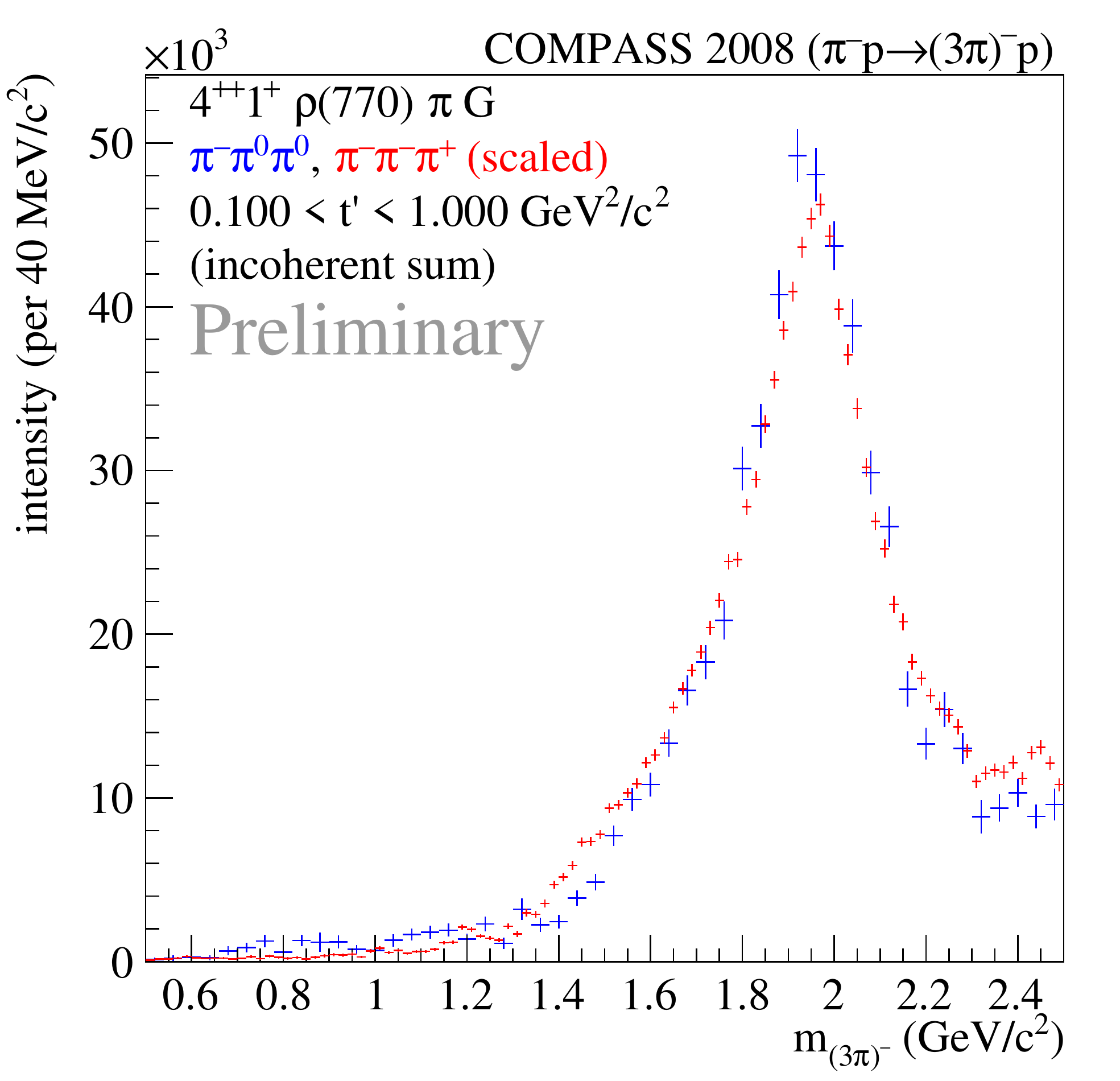} }
    \end{center}
  \end{minipage}
  \begin{minipage}[h]{.32\textwidth}
    \begin{center}
      \vspace{-0.1cm}
\resizebox{1.0\columnwidth}{!}{%
     \includegraphics[clip,trim= 5 0 10 15, width=1.0\linewidth, angle=0]{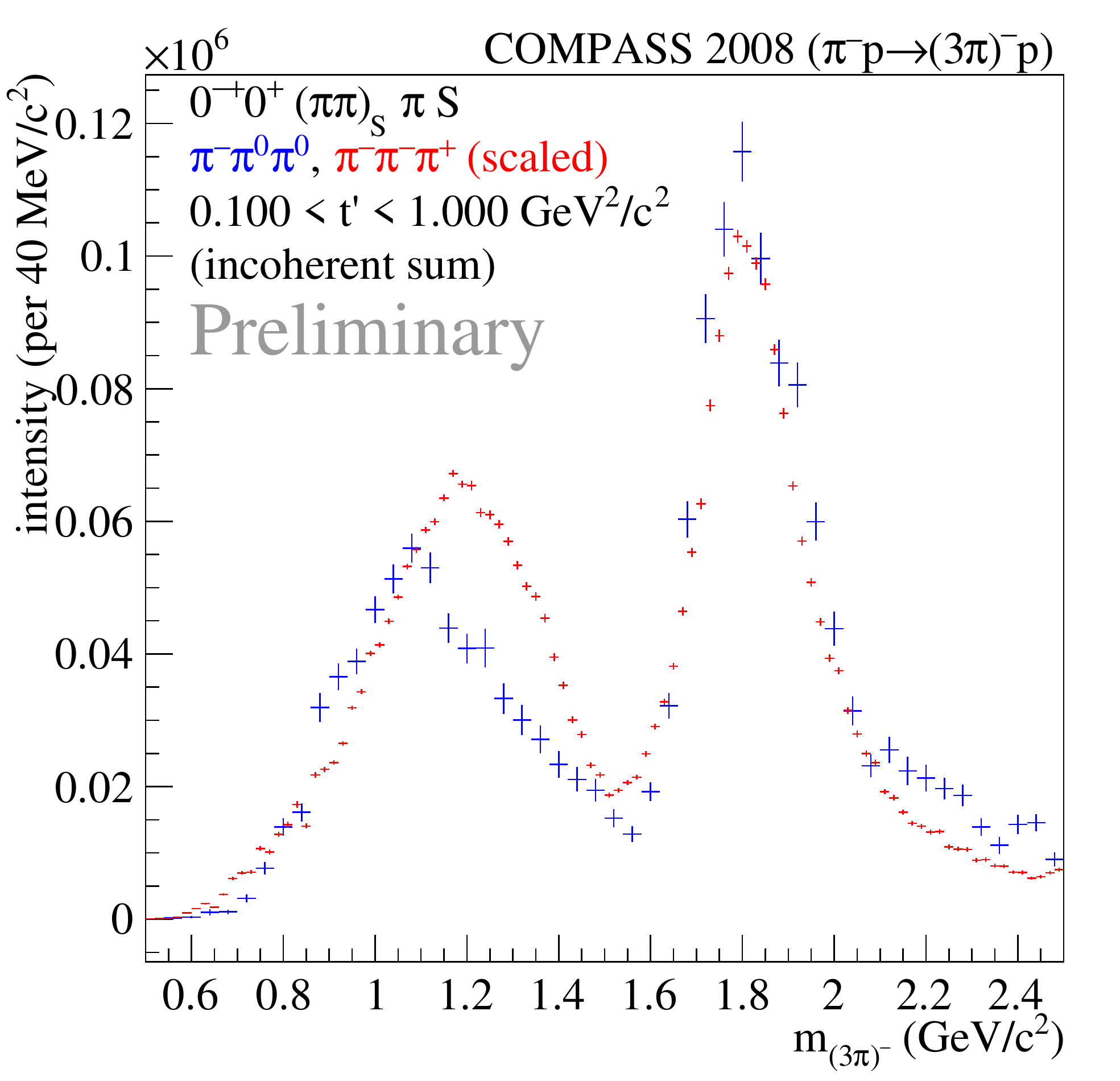}}                                                                 
    \end{center}
  \end{minipage}
    \begin{center}
      \vspace{-0.7cm}
     \caption{Incoherent sum over the different $t'$ ranges of mass-independently fitted intensities of major and smaller waves compared
       for neutral (blue) and charged (red) mode data. Shown are the three major waves (cf. also Fig.\,\ref{fig:MajorWavesTprimeDep}) 
       {\it(top)} and those, in which the $\pi(1800)$ {\it (bottom, left and right)} and the $a_4(2040)$ {\it (bottom, centre)} 
       are observed.  
}
       \label{fig:IncoherentSums}
     \end{center}
     \vspace{-0.7cm}
\end{figure}

Summing incoherently up the mass-independently fitted intensities of all $t'$ bins, one obtains a similar 
result as obtained previously~\cite{nerling:2012b}, where the mass-independent PWA were performed in bins of $m_{\rm 3\pi}$ only. 
The intensity spectra obtained with the extended method discussed here are basically smoother, as the $t'$ dependence is taken 
into account directly from the data instead of using the $t'$ slopes for the various resonances parameterised from the data.
For the three main waves, in which we observe the $a_1(1260)$, the $a_2(1320)$ and the $\pi_2(1670)$, shown for different 
$t'$ ranges in Fig.\,\ref{fig:MajorWavesTprimeDep}, the incoherent sums of intensities over all $t'$ ranges are shown in 
Fig.\,\ref{fig:IncoherentSums} (top) for each wave. Further incoherent sums are presented for the $(0^{-+})0^+f_0(980)\pi~S$-wave 
and the $(4^{++})1^+\rho(770)\pi~S$-wave (Fig.\,\ref{fig:IncoherentSums}, bottom, left and centre), 
showing the $\pi(1800)$ and $a_4(2040)$, respectively, and the $(0^{-+})0^+(\pi\pi)_{\rm s}\pi~S$-wave (Fig.\,\ref{fig:IncoherentSums}, 
bottom, right), where the peak at around 1.8\,GeV/$c^2$ can be attributed to the $\pi(1800)$. Other experiments claimed also a 
$\pi(1300)$ --- the object at the mass between 1 and 1.4\,GeV/$c^2$ in the $(0^{-+})0^+(\pi\pi)_{\rm s}\pi~S$-wave shows some differences 
for the neutral vs. the charged mode COMPASS data.
%
%
\clearpage
%
%
\begin{figure}[tp!]
    \begin{center}
         \vspace{-0.5cm}
\resizebox{1.0\columnwidth}{!}{%
  \includegraphics[clip,trim= 0 0 0 0, width=1.0\linewidth, angle=0]{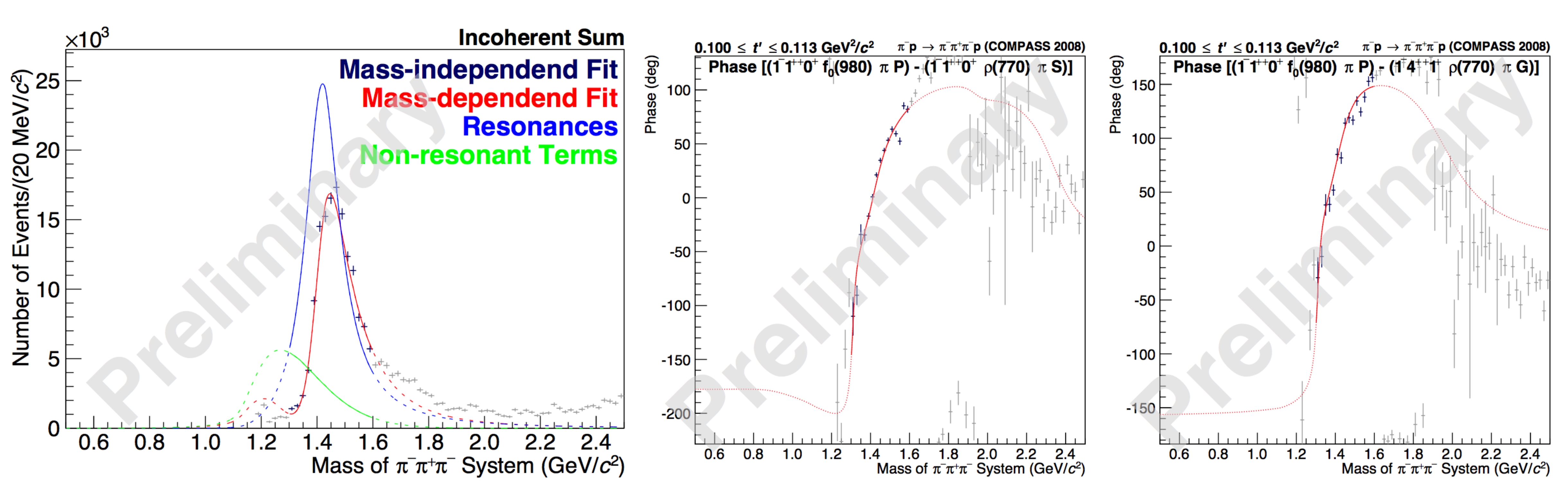} }
      \vspace{-0.7cm}
     \caption{PWA result for the $(1^{++})0^+f_0(980)\pi~P$-wave. 
       The mass-independently fitted intensities 
       for the neutral and the charged mode data show a narrow structure at 
       1420\,MeV/$c^2$
       (first presented in~\cite{suhl:2013} and \cite{paul:2013}). 
       For the charged mode data, the mass-independent fit result is shown
       (data points) with the 
       fitted BW description (red curve) overlaid, which consists of the BW describing the object at 1420\,MeV/$c^2$ 
       (blue curve) and the non-resonant background contribution (green curve) {\it (left)}. The corresponding 
       relative phases w.r.t.
       two other waves 
       are shown, again the mass-independent 
       result (data points) with the mass-dependent fit result overlaid (black line) {\it (centre and right)}. 
     }   
       \label{fig:a1-1420}
     \end{center}
     \vspace{-0.7cm}
\end{figure}
The PWA results for the $(1^{++})0^+f_0(980)\pi~P$-wave in the charged mode data are displayed in 
Fig.\,\ref{fig:a1-1420}.
The mass-independently fitted intensities (incoherent sums of the different $t'$ ranges) are shown 
(Fig.\,\ref{fig:a1-1420}, left).
The data exhibits, even though low in intensity, a strong enhancement in 
form of a clean narrow peak at 1420\,MeV/$c^2$, that is consistently observed for the neutral and the charged 
mode data, see also~\cite{suhl:2013} and \cite{paul:2013}, respectively.
The mass-independent PWA completed by the simultaneous BW fit to the result obtained from the mass-independent 
fit performed in the different ranges of $t'$ is shown for the $(1^{++})0^+f_0(980)\pi~P$-wave in the charged mode 
data in Fig.\,\ref{fig:a1-1420} (left). 
Here, the result of the mass-dependent fit is overlaid (curves) in addition. It consists of a 
BW describing the narrow object and some non-resonant component. No such object has previously been observed. 

The relative phases of this object observed in the $(1^{++})0^+f_0(980)\pi~P$-wave against the $a_1(1260)$ 
in the $(1^{++})0^{+}\rho(770)\pi~S$-wave and the $a_4(2040)$ in the $(4^{++})1^+\rho(770)\pi~G$-wave are shown 
in Fig.\,\ref{fig:a1-1420} (centre and right),
again with the mass-dependent fit result overlaid. 
A clean phase motion is observed exactly in the mass region of about 
1.3-1.6\,GeV/$c^2$, where the object is observed in the intensity plots, convincing that the object is of resonant 
nature. 

The neutral and the charged mode results 
are consistent not only for the well and less established 
states, but also for this new iso-vector resonance, that we call $a_1(1420)$. 
The new $a_1(1420)$ decaying into $f_0(980)\pi$ (and not observed in $\rho\pi$) is produced at a rather low intensity 
(less than 0.25\,\% of the total intensity), which might hint to a large strangeness content and an exotic nature. 
Similarly to $a_0(980)$ and $f_0(980)$, it might be the isospin-1 partner of the $f_1(1420)$ (that we observe well 
resolved decaying to $K\bar{K}\pi$~\cite{nerling:2011}) strongly coupling to $KK^*$, at least the similarity in width 
and mass is striking.  

\paragraph{Conclusions \& Outlook}~\\
In summary, we extended our PWA method to disentangle contributions from resonant and non-resonant production.
The neutral and the charged mode $(3\pi)^{-}$ data diffractively produced on a proton target show consistent results concerning 
major and less known resonances. In particular, we observe a new possibly exotic iso-vector state $a_1(1420)$ decaying into 
$f_0(980)\pi$ (and not observed in $\rho\pi$), having a width of about 140\,MeV/$c^2$ and showing resonant behaviour. 
The analyses will be proceeded, including the extraction of the $(\pi\pi)_s$ wave subsystem (not discussed here), before stronger 
conclusions will be drawn also on the disputed $\pi_1(1600)$ resonance, for which the interpretation as a narrow 
($\Gamma$=150-200\,MeV) resonance is excluded, and thus its nature can (partly) be connected to dynamical effects on top of a 
non-resonating Deck-like amplitude. 


\begin{thebibliography}{99} 
\bibitem{MeyerHaarlem2010} C.A.~Meyer and Y.Van.~Haarlem, Phys. Rev. C {\bf 82} (2010) {025208}.
\bibitem{Adams:1998} G.~S.~Adams {\it et al.}, Phys. Rev. Lett. {\bf 81}, (1998) 5760.
\bibitem{Khokhlov:2000} Y.~Khokhlov, Nucl. Phys. {\bf A663} (2000) 596.
\bibitem{Amelin:2005} D.~V.~Amelin {\it et al.}, Phys. Atom. Nucl. {\bf 68} (2005) 359.
\bibitem{Dzierba:2006} A.R.~Dzierba {\it et al.}, Phys. Rev. D {\bf 73} (2006) {072001}.
\bibitem{PDG} J.~Beringer {\it et al.}, (Particle Data Group), {Phys. Rev.} D {\bf 86} (2012) {010001}. 
\bibitem{Alekseev:2009a} M.~Alekseev {\it et al.}, COMPASS collaboration, Phys. Rev. Lett, {\bf 104} (2010) {241803}.
\bibitem{nerling:2009} F.~Nerling, AIP Conf. Proc. {\bf 1257} (2010) 286; arXiv:1007.2951[hep-ex].
\bibitem{nerling:2012b} F.~Nerling, EPJ Web Conf. {\bf 37} (2012) 09025; arXiv:1208.0474[hep-ex].  
\bibitem{haas:2011} F.~Haas, Conf. Proc., Hadron2011, Munich (2011); arXiv:1109.1789v2[hep-ex].
\bibitem{Accmor:1980} C.~Daum {\it et al.}, Phys. Lett. {\bf 89}B (1980) {276}; Phys. Lett. {\bf 89}B (1980) {281}.  
\bibitem{Deck:1964} R.T.~Deck, Phys. Rev. Lett. {\bf 13} (1964) {169}.  
\bibitem{paul:2013} S.~Paul, Conf. Proc., MENU 2013, Rom, Italy, (2013); arXiv:1312.3678[hep-ex].
\bibitem{suhl:2013} S.~Uhl, Conf. Proc., Hadron2013, Nara, Japan, (2013); arXiv:1401.4943[hep-ex].
\bibitem{nerling:2011} J. Bernhard and F. Nerling, Conf. Proc. Hadron2011, Munich (2011); \\arXiv:1109.0219[hep-ex].
\end{thebibliography}
\end{document}